\documentclass[fleqn,10pt]{wlscirep}
\usepackage[pdftex]{dropping}
\usepackage{subfigure}
\usepackage{bookmark}

\title{The effect of oscillator and dipole-dipole interaction on multiple optomechanically induced transparency in cavity optomechanical system}

\author[1]{Jin-Lou~Ma}
\author[1,+]{Lei~Tan}
\author[1]{Qing~Li}
\author[2]{Huai-Qiang~Gu}
\author[3]{Wu-Ming~Liu}

\affil[1]{Institute of Theoretical Physics, Lanzhou University, Lanzhou $730000$, China}
\affil[2]{School of Nuclear Science and Technology, Lanzhou University, Lanzhou $730000$, China}
\affil[3]{Beijing National Laboratory for Condensed Matter Physics, Institute of Physics, Chinese Academy of Sciences, Beijing 100190, China}
\affil[+]{Correspondence and requests for materials should be addressed to L.T.(email:tanlei@lzu.edu.cn)}

\begin{abstract}
 We theoretically investigate the optomechanically induced transparency (OMIT) phenomenon in a $N$-cavity optomechanical system doped with a pair of Rydberg atoms with the presence of a strong control field and a weak probe field applied to the $N$th cavity. It is found that $2N\!-\!1(N<10)$ numbers of OMIT windows can be observed in the output field when $N$ cavities couple with $N$ mechanical oscillators and the mechanical oscillators coupled with different even- or odd-labelled cavities can lead to diverse effects on OMIT. Furthermore, the ATS effect appears with the increase of the effective optomechanical coupling rate. On the other hand, two additional transparent windows (extra resonances) occur, when two Rydberg atoms are coupled with the cavity field. With DDI strength increasing, the extra resonances move to the far off-resonant regime but the left one moves slowly than the right one due to the positive detuning effect of DDI. During this process, Fano resonance also emerges in the absorption profile of output field. \newline
\end{abstract}

\begin{document}

\flushbottom

\maketitle

\thispagestyle{empty}

In atomic systems, electromagnetically induced transparency (EIT)\cite{Harris1990,Boller,Fleischhauer} is induced by quantum interference effects or Fano-interactions\cite{Fano} due to the coherently driving
atomic wavepacket with an external control laser field. The OMIT, a phenomenon analogous to the EIT, was predicted theoretically firstly\cite{Agarwal,Huang} and then verified experimentally\cite{Weis,Safavi} in a cavity optomechanical system which is caused by the destructive quantum interference between different pathways of the internal fields. More recently, the study of OMIT has attracted much attention. For instance, the single-photon routers\cite{Agarwal1}, the ultraslow light propagation\cite{Teufel}, the quantum ground state cooling\cite{Liu}, the precision measurement\cite{Zhang}, the Brillouin scattering induced transparency and non-reciprocal light storage\cite{Dong,Kim}, the optomechanically induced amplifcation\cite{Yan}, the effective mass sensing\cite{Gao},
control of photon propagation in lossless media\cite{LHe}, optomechanically induced stochastic resonance\cite{Monifi} and chaos transfer and the parity-time-symmetric microresonators\cite{Jing}. In addition, tunable EIT and absorption\cite{HIan}, polariton states\cite{xgu} and transition from blockade to transparency\cite{yxliu} in a circuit-QED system have also been studied. On the other hand, the studies on the OMIT have been extended to  double- and multi-optomechanically induced transparency\cite{zmzhang} by integrating more optical or mechanical modes. It has been reported that multiple OMIT windows may occur in the atomic-media assisted optomechanical system\cite{Xiao,Wang,Akram}, multi-resonators optomechanical system\cite{Huang2014}, optomechanical system with N membranes\cite{Huang1403}, two coupled optomechanical systems\cite{Sohail2017} and the multi-cavity optomechanical system\cite{Sohail}. In particular, achieving multi-OMIT phenomenon shows many practical applications for the multi-channel optical communication and quantum information processing, which motivate the further investigation on such OMIT.

Currently, a hybrid cavity optomechanical system containing atoms has attracted much attention. The additional control of atomic freedom can lead to rich physics resulted from the enhanced nonlinearities and the strengthened coupling strength, which can also provide an coherent optical controlled method to change the width of the transparency window\cite{Ian,Akram,Xiao}, multistability of OMIT\cite{chang} and switch from single to double and multiple OMIT windows\cite{Sohail}. On the other hand, there has been a great interest in studying the phenomenon of EIT in the interacting Rydberg atoms system due to  the strongly long range dipole-dipole interactions (DDI) or van der Waals interactions and long radiative lifetimes for many years\cite{Fleischhauer}. Based on the essential blockade effect arising from DDI, some novel behaviors in EIT are revealed, such as the transmission reduction\cite{Petrosyan},  the nonlocal propagation and enhanced  absorption\cite{Li}, the nonlocal Rydberg EIT\cite{Wu}, the nonlinear Rydberg-EIT\cite{Liu2014}, and the dipolar exchange induced transparency\cite{Petrosyan2017}. Furthermore, optomechanical cavity system assisted by Rydberg atomic ensembles has been proposed to investigate the state transfer, sympathetic cooling and the non-classical state preparation\cite{Guerlin,Carmele}. It can also be found that an all-optical transistor can be manipulated by controlling the Rydberg excitation\cite{Liu1}. Even though many meaningful researches of EIT based
on Rydberg atoms have been conducted, further studies on OMIT with the auxiliary DDI Rydberg atoms are also expected.

Motivated by the remarkable developments and potential applications in OMIT mentioned above, in the present work we will study the multiple OMIT in a multi-cavity optomechanical system (MCOS) assisted by a pair of DDI Rydberg atoms driven by two coupling fields. Different from the  previous studies, we focus on a multi-cavity optomechanical system composed of $N$ optical modes and $N$ mechanical modes. The Heisenberg-Langevin equations for the hybrid MCOS are solved and the in-phase and out-of-phase quadratures of the output field based on the the input-output theory are obtained to determine the effects  of the odd and even labelled  oscillators and DDI on the multi-OMIT. It can be found that the multi-OMIT and Fano resonance can be controlled by the DDI.

The paper is organized as follows: In Sec. \uppercase\expandafter{\romannumeral2}, we introduce the multi-cavity optomechanical system and the Hamiltonian of our system, and Sec. \uppercase\expandafter{\romannumeral3} is devoted to obtaining
 the Langevin Equations of the system and the output field based on the input-output theory. The effects of the mechanical oscillators and DDI on OMIT are discussed in Sec. \uppercase\expandafter{\romannumeral4}. Finally the conclusions are summarized in Sec. \uppercase\expandafter{\romannumeral5}.

\section*{{\protect\LARGE \textbf{Results}}}

\subsection{Theoretical model and Hamiltonian.}

\begin{figure}[!htb]
\centerline{\includegraphics[width=0.5\textwidth]{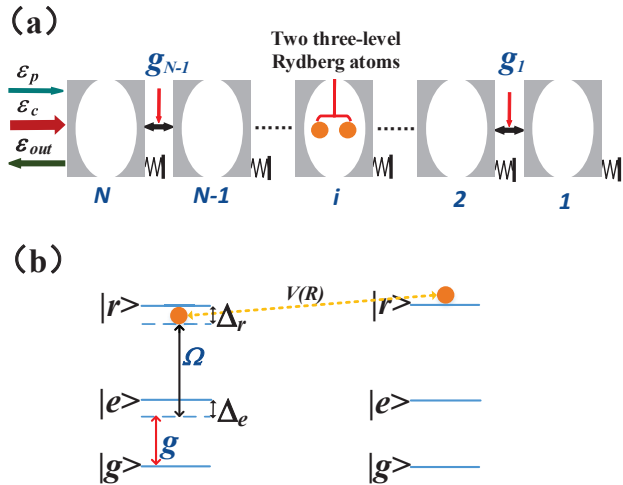}}
\caption{Schematic diagram of the multi-cavity optomechanical system.
(a) $N$ cavities connect through hopping rates $g_n$. A pair of Rydberg atoms are put into the $i$th cavity.
(b) The pair of ladder-type three-level Rydberg atoms interact with each other and one of Rydberg atoms is excited in the Rydberg state during the process of interaction. $g$ and $\Delta_e$ are the coupling strength and the frequency detuning of the transition $|g\rangle \leftrightarrow |e\rangle$, respectively. $\Omega$ and $\Delta_r$ are the Rabi frequency and the frequency detuning of the transition $|e\rangle \leftrightarrow |r\rangle$, respectively. In addition, $V(R)$ is the DDI strength between two Rydberg atoms.}
\label{fig1}
\end{figure}

The $1D$ MCOS under consideration is shown in Fig.~\ref{fig1}. The $Nth$ cavity of the cavity optomechanical arrays is coherently driven by a strong control field of frequency $\omega_c$ and a weak probe laser field of frequency $\omega_p$.
$N$ optomechanical cavities are labelled as $1$ , $2$ , $\cdots $, $N$. The frequencies of $jth$ cavity and $jth$ mechanical oscillator are denoted by $\omega_j$ and $\omega_{mj}$, respectively. The coupling strength between $jth$ cavity and $jth$ mechanical oscillator is $g_{mj}$, and $g_n$ is the hopping rate between $n$th and $(n+1)th$ cavities $(n\not=N)$. In addition, a pair of DDI ladder-type three level Rydberg atoms are assisted in the $ith$ cavity.
The Rydberg atoms of our system may chose Cesium (Cs) atoms, the fine-structure states $|6S_{1/2}, F=4\rangle$ and $|6P_{3/2}, F^{\prime}=5\rangle$ can be regarded as the ground state $|g\rangle$ and the intermediate state $|e\rangle$, respectively, while the correspond Rydberg state $|r\rangle$ is assumed as $^{70}S_{1/2}$ \cite{Goban}.
As for the first Rydberg atom, the frequency of control field $\omega_c$ is coupled to the $|e\rangle \leftrightarrow |r\rangle$ transition with a Rabi frequency $\Omega$ and a frequency detuning $\Delta_r$. The $ith$ cavity field drives the $|g\rangle \leftrightarrow |e\rangle$ transition with strength $g$ and the frequency detuning $\Delta_e$. In brief, the second Rydberg atom is assumed to be excited in the Rydberg state and coupled with the first Rydberg atom by DDI in the $ith$ cavity ($1\le i\le N$) due to the long lifetime ($\tau\geq100\mu s$) of the Rydberg state. As explained in Refs.\cite{Peyronel,Beguin,Neuzner}, this configuration has an experimental feasibility when the radius of the blockade is smaller than the interatomic distance of a pair of Rydberg atoms, then they can be excited to the Rydberg state simultaneously and their interactions are utilized via van-der-Waals type of DDI.

The total Hamiltonian $H$ of the hybrid cavity optomechanical system in
the rotating-wave frame can be written as
\begin{equation}
H = H_c + H_m + H_a +H_{in}+ H_{int},\label{H1}
\end{equation}
where the first four terms describe the Hamitonians of the optical cavity, the mechanical oscillator, the two Rydberg atoms and the input fields, with the expressions as following
\begin{align}
\begin{split}
H_c =& \sum_{j=1}^{N}\Delta_{j}c^{\dagger}_{j}c_{j},\\
H_m =& \sum_{j=1}^{N}\omega_{mj}b_j^{\dagger}b_j,\\
H_a =& \Delta_{e}\sigma_{ee}^{(1)} +(\Delta_{e}+\Delta_{r})\sigma^{(1)}_{rr}+ \omega_{rg}\sigma^{(2)}_{rr},\\
H_{in} =& i\varepsilon_{c}(c_{_{N}}^{\dagger}-c_{_{N}})+i\varepsilon_{p}(c_{_{N}}^{\dagger}e^{-i\Delta t}-c_{_{N}} e^{i\Delta t}).
\end{split}
\label{H2}
\end{align}

The optical modes are described as an annihilation (creation) operator $c_j$($c_{j}^\dagger$) of the $jth$ cavity field, and $b_j^\dagger$($b_j$) is the creation (annihilation) operator of the $jth$ mechanical resonator. $\Delta_j= \omega_j-\omega_c$ is the detuning of the $jth$ cavity field from the control field, and  $\Delta=\omega_p-\omega_c$ represents the detuning between the probe field and the control field. $\Delta_{e} = \omega_{eg}-\omega_j$, $\Delta_{r} = \omega_{re}-\omega_p$, and $\omega_{\mu \nu}$ represents the frequency of the atomic transition between the level $|\mu\rangle$ and level
 $|\nu\rangle (\mu,\nu=g,e,r)$. $\sigma_{\mu\nu}^{(k)}\equiv|\mu\rangle_{kk}\langle\nu|$ is the projection $(\mu=\nu)$ or
 transition $(\mu\ne\nu)$ operator of the $kth$ ($k=1,2$) Rydberg atom. Moreover, the Hamiltonian of the input fields includes the Hamiltonian of the control field and probe field. $\varepsilon_c$ is the control field amplitude and $\varepsilon_p$ is the probe field amplitude.

The last term of Eq. (\ref{H1}) describes the system's interaction Hamiltonian,
\begin{align}
H_{int}=& \sum_{n=1}^{N-1}g_{n}(c_{n+1}^{\dagger}c_{n}+c_{n+1}c_{n}^{\dagger})-
\sum_{j=1}^{N}g_{mj}(c_{j}^{\dagger}c_{_j})(b_j^{\dagger}+b_j)+(\Omega \sigma_{er}^{(1)}+g c_i\sigma_{eg}^{(1)} +H.c)+V(R)\sigma_{rr}^{{(1)}}\sigma_{rr}^{{(2)}}.
\label{H3}
\end{align}
In Eq. (\ref{H3}), the first term corresponds to the hopping between the two adjacent cavities and $g_n$ is the intercavity tunneling strength. The second term describes the interaction between the $jth$ cavity and the mechanical oscillator via the radiation pressure and $g_{mj}$ is the coupling strength. One of the Rydberg atoms
interacted with the control field and $ith$ cavity field is listed in the third term, respectively. $V(R)$ is the DDI strength between two Rydberg atoms which is described as the last term, and $R$ is the distance between two Rydberg atoms which can be controlled at different ranges by the separate optical traps\cite{Beguin}.

\subsection{The dynamical equation.}
The Heisenberg-Langevin equatons for the operators can be obtained based on the Hamiltonian (\ref{H1}). Using the the factorization assumption (mean field approximation), viz, $\langle QC\rangle=\langle Q\rangle\langle C\rangle$\cite{Agarwal,Akram2015}, the equations of the mean value of the operators can be given by
 \begin{align}
 \begin{split}
   \langle\dot{c}_{_N}\rangle =&-(\kappa_{_N}+i\tilde\Delta_{_N})\langle c_{_N}\rangle-ig_{_{N-1}}\langle c_{_{N-1}}\rangle+\varepsilon_c+\varepsilon_pe^{-i\Delta t}
   + ig_{mN}\langle c_N\rangle(\langle b_N^{\dagger}\rangle+\langle b_N\rangle),\\
   \langle\dot{c}_n\rangle =&-(\kappa_n+i\tilde\Delta_n)\langle c_n\rangle-i(g_{n-1}\langle c_{n-1}\rangle+g_{n}\langle
   c_{n+1}\rangle)+ ig_{mn}\langle c_n\rangle(\langle b_n^{\dagger}\rangle+\langle b_n\rangle),n\neq1,i,N,\\
 \langle\dot{c}_1\rangle =&-(\kappa_1+i\tilde{\Delta}_1)\langle c_1\rangle-ig_{1}\langle c_{2}\rangle+ig_{m1}\langle c_1\rangle(\langle b_1^{\dagger}\rangle+\langle b_1\rangle),\\
 \langle\dot{b}_j\rangle =&-(\gamma_{mj}+i\omega_{mj})\langle b_j\rangle+ig_{mj}|\langle c_{j}\rangle|^2,\\
\langle\dot{\sigma}_{ge}\rangle =& -(\gamma_{ge}+i\Delta_{e})\langle\sigma_{ge}\rangle+ig(\langle\sigma_{ee}\rangle-\langle\sigma_{gg}\rangle)\langle c_{i}\rangle-i\Omega
\langle{\sigma}_{gr}\rangle,\\
   \langle\dot{\sigma}_{gr}\rangle=&-(\gamma_{gr}+iS+i\Delta_{r})\langle{\sigma}_{gr}\rangle+ig\langle{\sigma}_{er}\rangle\langle c_i\rangle-i\Omega\langle\sigma_{ge}\rangle,\\
\langle\dot{\sigma}_{er}\rangle =&-(\gamma_{er}+i\Delta_{r}+iS-i\Delta_{e})\langle{\sigma}_{er}\rangle+ig\langle{\sigma}_{gr}\rangle\langle c_i\rangle +i\Omega(\langle\sigma_{rr}\rangle-\langle\sigma_{ee}\rangle),
\end{split}
\label{H4}
\end{align}
For two Rydberg atoms trapped in the $ith$ cavity case
\begin{align}
\langle\dot{c}_i\rangle =&-(\kappa_i+i\tilde\Delta_i)\langle c_i\rangle-i(g_{i-1}\langle c_{i-1}\rangle+g_{i}\langle c_{i+1}\rangle)
  -ig\langle\sigma_{ge}\rangle +ig_{mi}\langle c_i\rangle(\langle b_i^{\dagger}\rangle+\langle b_i\rangle),i\neq1,N. \label{H5}
   \end{align}
If two Rydberg atoms are confined in the first cavity, Eq.~(\ref{H5}) should be substituted by
 \begin{align}
\langle\dot{c}_1\rangle =&-(\kappa_1+i\tilde{\Delta}_1)\langle c_1\rangle-ig_{1}\langle c_{2}\rangle+ig_{m1}\langle c_1\rangle(\langle b_1^{\dagger}\rangle+\langle b_1\rangle)-ig\langle\sigma_{ge}\rangle. \label{H6}
 \end{align}
If one puts the Rydberg atoms into the $Nth$ cavity, Eq.~(\ref{H5}) should be replaced by
 \begin{align}
 \langle\dot{c}_{_N}\rangle =&-(\kappa_{_N}+i\tilde\Delta_{_N})\langle c_{_N}\rangle-ig_{_{N-1}}\langle c_{_{N-1}}\rangle+\varepsilon_c+\varepsilon_pe^{-i\Delta t}
-ig\langle\sigma_{ge}\rangle+ig_{mN}\langle c_N\rangle(\langle b_N^{\dagger}\rangle+\langle b_N\rangle), \label{H7}
 \end{align}
where $\kappa_j$ and $\gamma_{mj}$ are introduced phenomenologically to denote the dissipation of the $jth$ cavity, and the decay rate of the $jth$ mechanical oscillator, respectively. $S=V(R)$ with $\bar\sigma_{rr}^{(2)}=1$ due to reason that the second Rydberg atom are assumed to be excited to the Rydberg state during the interaction process with the first Rydberg atom. $\gamma_{\mu\nu}(\mu,\nu=g,e,r)$
is the decay rate of transition between the level $|\mu\rangle$
and the level $|\nu\rangle$.  In addition, $\tilde\Delta_{j}=\Delta_{j}-g_{mj}\bar\lambda_j$. The general form of $\bar\lambda_j$ will be given in the following.
In order to obtain the steady-state solutions, which are exacted for the control field
in the parameter $\varepsilon_c$ and corrected to the first order in the parameter
$\varepsilon_p$ of the probe field. As the probe field is much weaker
than the control field, then the average value of the operator ${O}$ can be approximately written by using the ansatz\cite{Boyd}
  \begin{equation}
    \langle{O}\rangle\!=\!\bar{O}+\delta O(t)=\bar{O}+O_{-} e^{-i\Delta t}+O_{+} e^{i\Delta t}. \label{H8}
  \end{equation}
where $\bar{O}$ describes the steady-state value of the operator ${O}$ governed by the control field, but $\delta O(t)$ is proportional to the weak probing field, which gives rise to the Stokes scattering and the anti-Stokes scattering of light from the strong control field.  Subsequently, substituting Eq.~(\ref{H8}) into Eqs.~(\ref{H4})-(\ref{H7}), one can obtain the steady-state solutions of the Heisenberg-Langevin equations. Because $\bar O$ is independent of time, and $\delta O(t)$ of the same order as $\varepsilon_p$ depends on the time but remains much smaller than $\bar O$, one can separate the equations into two parts. One part is irrelevant of time and the other one is related to the time. Assuming that the cavity optomechanical system\cite{LiaoJQ,LiaoJQ1,LiaoJQ2,LiaoJQ3,LiaoJQ4} evolves in the resolved sideband regime, e.g., $\kappa_j\ll\omega_{mj}$, then the Stokes part, the low sidebands and off-resonant one, can be ignored i.e., $O_+\approx 0$ in Eq.~(\ref{H8}), only the anti-Stokes scattering survives in the hybrid system. Thus, all elements of $O_-$ can be obtained as follows using the above ansatz,
\begin{align}
\begin{split}
 0 =&-(\kappa_{_N}-ix_N)c_{_N,-}-ig_{_{N-1}} c_{_{N-1,-}}+\varepsilon_pe^{-i\Delta t}+iG_{mN} b_{N,-},\\
 0 =&-(\kappa_n-ix_{n})c_{n,-}-i(g_{n-1} c_{n-1,-}+g_{n}c_{n+1,-})+ iG_{mn} b_{n,-},n\ne 1,i,N,\\
 0 =&-(\kappa_i-ix_i) c_{i,-} -i(g_{i-1}c_{i-1,-} +g_{i}c_{i+1,-} )+ iG_{mi} b_{i,-}-ig\sigma_{ge,-},i\ne 1,N,\\
 0 =&-(\gamma_{ge}-ix)\sigma_{ge,-} +ig(\bar\sigma_{ee} -\bar\sigma_{gg} )c_{i,-} -i\Omega\sigma_{gr,-},\\
 0 =&-(\gamma_{gr}-ix_{gr})\sigma_{gr,-}+ig(\sigma_{er,-} \bar{c}_i+\bar{\sigma}_{er}c_{i,-} )-i\Omega\sigma_{ge,-},\\
 0 =&-(\gamma_{er}-ix_{er})\sigma_{er,-} +ig(\sigma_{gr,-}\bar{c}_i +\bar{\sigma}_{gr}c_{i,-})+ i\Omega(\bar\sigma_{rr} -\bar\sigma_{ee} ),\\
 0 =&-(\kappa_1-ix_1)c_{1,-} -ig_{1}c_{2,-} +iG_{m1} b_{1,-},\\
 0 =&-(\gamma_{mj}-ix_j)b_{j,-} +iG_{mj}^* c_{j,-},\\
\end{split}
\label{H9}
\end{align}
As we provide the equations in the resolved sideband regime, the detuning parameters are set as $\tilde\Delta_{j}=\Delta_{j}=\Delta_{r}=\Delta_{e}=\omega_{mj}$, with $x_{er}=\Delta-\Delta_r-S$ and $x_{gr}=\Delta-\Delta_r-\Delta_e-S$. $x_j=\Delta-\omega_{mj}$ is the detuning from the center line of the sideband. $G_{mj}^*=g_{mj}\bar c_j^*$ and $G_{mj}=g_{mj}\bar c_j$ describe the effective optomechanical coupling rate of the $jth$ cavity and they are equal. By solving the equations for $\bar{O}$ of the mechanical oscillators, one can obtain
 \begin{equation}
\bar\lambda_j\equiv \bar{b}_j+\bar{b}_j^*=\frac{2\omega_{mj} g_{mj}|\bar c_j|^2}{\gamma_{mj}^2+\omega_{mj}^2}.\label{H10}
 \end{equation}

\subsection{The output field.}
 The response of the system  can be detected by the output field at the probe frequency, which can be expressed as follows via the standard input-output theory of the cavity\cite{Walls},
\begin{equation}
 \varepsilon_{out,p}e^{-i\Delta t}+\varepsilon_p e^{-i\Delta t}+\varepsilon_c =2\kappa_{_{N}}\langle
 c_{_N}\rangle.\label{H11}
\end{equation}
Therefore, one can express the total output field as
\begin{equation}
  \varepsilon_T=\frac{\varepsilon_{out,p}}{\varepsilon_p}+1=\frac{2\kappa_{_N}
  c_{_{N,-}}}{\varepsilon_p}=\chi_p+i\tilde{\chi}_p.\label{H12}
\end{equation}
Here, $\chi_p=Re(\varepsilon_T)$ and $\tilde{\chi}_p=Im(\varepsilon_T)$ denote the in-phase and out-of-phase quadratures of the output field associated with the absorption and dispersion, respectively. The OMIT is the phenomenon of the simultaneously vanishing absorption and dispersion. These two quadratures of the output field can be measured via the homodyne technique\cite{Walls}. Using Eq.~(\ref{H9}), $c_{N,-}$ can be easily obtained, then the expression of the output field $\varepsilon_T$ is given in a constructive form,
\begin{eqnarray}
\varepsilon_T\!=\!2\kappa_{_N}c_{_{N,-}}\!=\!\frac{2\kappa_{_N}}{B_N+\frac{g^2_{_{N-1}}}{B_{N-1}+\frac{g^2_{_{N-2}}}{\frac{\scriptstyle\ddots}{
B_i+A+\frac{g_{i-1}^{2}}{\frac{\scriptstyle\ddots}{{B_2+\frac{g^2_{1}}{B_1}}}}}}}},\label{H13}
\end{eqnarray}
where $B_j=\kappa_{j}-ix_j+\frac{|G_{mj}|^2}{\gamma_{_mj}-ix_j}(j=1,...,N)$. In the above equation, the first line of the denominator describes two cavities with decay rates $\kappa_{N}$ and $\kappa_{N-1}$ are coupled through the coupling strength $g_{N-1}$. Second line of the denominator describes the interaction of two cavities with decay rates $\kappa_{N-1}$ and $\kappa_{N-2}$ and the coupling strength is $g_{N-2}$ and so on. It is obvious that each line of the denominator contains an interaction term denoted by an effective coupling $G_{mj}$ between the mechanical oscillator and the cavity. Analytically, we note that when $G_{mj}=0$, the mechanical oscillator is not coupled with $jth$ cavity. Moreover, the extra term $A$ in the $B_i$ line represents the interaction of the cavity field with the pair of  Rydberg atoms including DDI, and its general form is shown in Eq.~(\ref{H14}) in the following with $Q=(\gamma_{gr}+i\Delta_{r}+iS)(\gamma_{ge}+i\Delta_{e})+\Omega^2$,
$P=i(\Delta_r+S-\Delta_e)+\gamma_{er}+\frac{Ge^2(\gamma_{ge}+i\Delta_{e})}{(\gamma_{gr}+i\Delta_{r}+iS)(\gamma_{ge}+i\Delta_{e})+\Omega^2}$, and $G_{e}=g\bar{c}_{i}$ is the effective coupling strength between the Rydberg atom and the cavity field. Certainly, when one traps the atoms in the first cavity, this term will appear in the last line. If Rydberg atoms are localized in the $Nth$ cavity, it will emerge in the first line of the denominator.
\begin{equation}
  A\!=\!\frac{[g^2(\gamma_{gr}-ix_{gr}+\frac{G_e^2}{\gamma_{er}-ix_{er}})+\frac{(g\Omega G_{e})^2}{PQ}](\bar\sigma_{rr}+2\bar\sigma_{gg}-1)
-\frac{(g\Omega)^2}{P}(2\bar\sigma_{rr}+\bar\sigma_{gg}-1)}
{(\gamma_e-ix_i)(\gamma_{rg}-ix_{rg}+\frac{G_e^2}{\gamma_{er}-ix_{er}})+\Omega^2}.\label{H14}
\end{equation}

From Eq.~(\ref{H14}), it can be found that the output field depends on $\bar c_{j}$ of the $jth$ cavity and the population $\bar\sigma_{gg}$ $(\bar\sigma_{rr})$ of the ground (Rydberg) state, which can be determined by solving Eq.~(\ref{H9}) for all $\bar O$. Note that there are four kinds of direct interactions in the system: the coupling between the adjacent cavities, the interaction between the cavities and the oscillators, the interactions of
the cavities with the Rydberg atoms and the DDI between the Rydberg atoms, which make the expressions of $\bar c_{j}$, $\bar\sigma_{gg}$ and $\bar\sigma_{rr}$ become very complicated, then it is too difficult to give concrete forms. Fortunately, the values of  $\bar c_{j}$ only affect the width of the OMIT windows\cite{Sohail}. When one focuses on the numbers of the OMIT window by numerical computation, $G_{mj}$ and $G_e$ can be valued by any reasonable and convenient value. Same argument, we also assume that the average $\bar\sigma_{gg}= 1$ and $\bar\sigma_{rr}= 0$. Besides, to  benefit more OMIT windows as many as possible, the system works in the weak dissipative regime, i.e, $g_j \geq \kappa_N\gg\kappa_j,\gamma_{mj/gr/er/ge}$.

Without loss of generality, it is assumed that the parameters of the system are chosen as follows. For the mechanical oscillator,  $\gamma_{m1}=\gamma_{m2}=\dots=\gamma_{mN}$, for the effective optomechanical rates, $G_{m1}=G_{m2}=\dots=G_{mN}$;
The cavity decay rates are $\kappa_1=\kappa_2=\dots=\kappa_{N-1}$, the tuneling parameters are set as $g_1=g_2=\dots=\kappa_N$, the frequencies of mechanical oscillators are $\omega_{m1}=\omega_{m2}=\dots=\omega_{mN}$, therefore, the detunings from the center line of the sidebands are the same $x_1=\dots=x_N\equiv x$. \\

\subsection{Without Rydberg atoms.}
 In this section, we first focus on the multiple OMIT phenomenon emerged due to the interaction between the cavity field and the mechanical oscillators without the Rydberg atoms. The parameters are  $\omega_{mN}/g_{mj}=20$, $\gamma_{mN}/g_{mj}=0.001$, $\kappa_{N-1}/g_{mj}=0.002$, $\kappa_{N}/g_{mj}=2$, and we assume $G_{mN}/g_{mj}=\bar c_j=1$. The optomechanical coupling parameter $g_{mj}=1$ kHz
is based on the realistic cavity optomechanical system\cite{Weis}. For simplicity, the following absorption analysis of the output field are restricted to a hybrid system with four cavities. The generalization to a large number of cavities case can be made according to the same method mentioned based on Eqs.~(\ref{H9})-(\ref{H14}).

\begin{figure}[t]
\centerline{\includegraphics[width=0.5\textwidth]{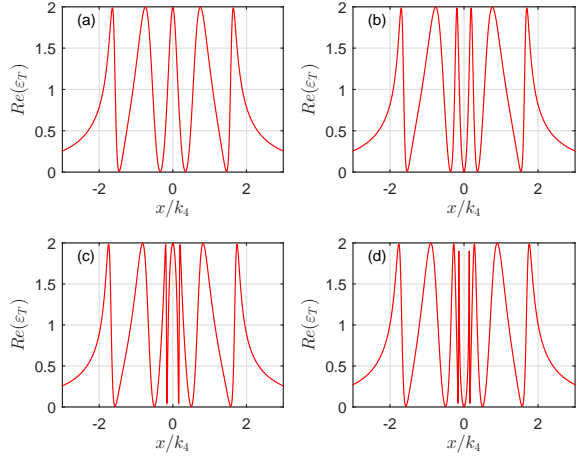}}
\caption{The absorption $Re(\varepsilon_T)$ as a function of $x/\kappa_4$ for four cavities. The subplot (a) corresponds to one mechanical oscillator coupled to cavity $1$, the subplot (b) describes two mechanical  oscillators coupled to cavity $1$ and $2$, respectively. The subplot (c) shows three mechanical oscillators coupled to cavity $1$, $2$ and $3$. The subplot (d) illustrates four mechanical oscillators coupled to cavity $1$, $2$, $3$ and $4$.}
\label{fig2}
\end{figure}

\begin{figure}[t]
\centerline{\includegraphics[width=0.90\textwidth]{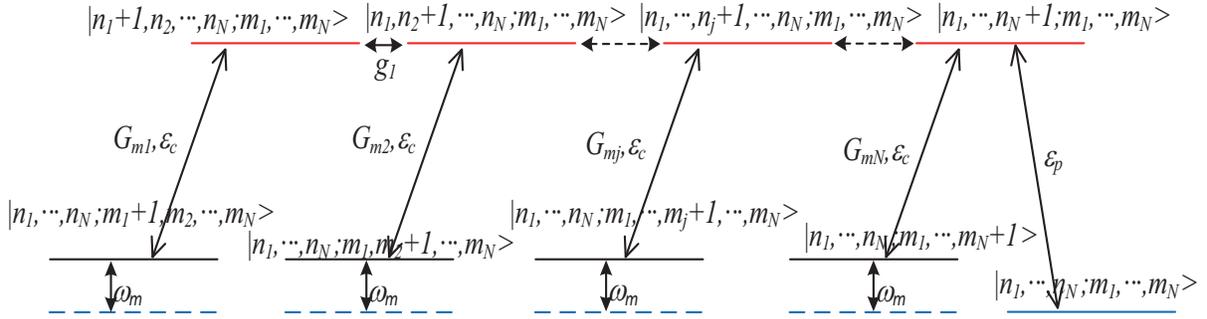}}
\caption{Energy level structure of the multi-cavity optomechanical system coupled with multi-oscillator.
The number state of photons and phonons are denoted by $n_j$ and $m_j$. The tunneling parameter between $|n_1,...,n_N;m_1,m_2,...m_N\rangle$ and $|n_1,n_{2}\!+\!1,...n_N;m_1,m_2,...m_N\rangle$ is $g_i$, the coupling strength between $|n_1,...,n_{i}\!+\!1,...n_N;m_1,...,m_i,...m_N\rangle$ and $|n_1,...,n_{i},...n_N;m_1,...,m_i\!+\!1,...m_N\rangle$ is $G_{mj}$.}
\label{fig3}
\end{figure}

Firstly, Fig.~\ref{fig2} illustrates the absorption $Re(\varepsilon_T)$ of the output field as a function of $x/\kappa_N$ for four cavities. In detail, Fig.~\ref{fig2}(a) describes only one mechanical oscillator coupled to the first cavity. The mechanical oscillators are coupled to the first and second cavities are shown in Fig.~\ref{fig2}(b). Fig.~\ref{fig2}(c) corresponds to three mechanical oscillators coupled to cavity $1$, $2$ and $3$, respectively. Fig.~\ref{fig2}(d) depicts four mechanical oscillators coupled to four cavities. The dips of the absorption line correspond to the  transparency windows of the output field. From Fig.~\ref{fig2}, it can be found that the number of transparency windows adds one with the increase of the mechanical oscillator in turn, which is determined by the infinity denominator of Eq.~(\ref{H13}) corresponding to the appearance of the coupling parameters $g_{N-1}$ and $G_{mN}$ in the denominators. When the hybrid system has $N$ cavities coupled with $N$ mechanical oscillators one by one without considering the effects of the outside environment, the sum of transparency windows adds to $2N\!-\!1$. Thus, MCOS becomes transparent to the probing field at $2N\!-\!1$ different frequencies, which are the destructive interferences between the input probing field and the anti-Stokes fields generated by the interactions of the coupling cavity field within the multiple cavities and the interactions between the coupling cavity field and the mechanical oscillators. However, when $N$ becomes large and each cavity  couples with its bath, numerical results show that the multiple transparency windows of this system become more and more opaque. Therefore, what we are concerned only the small $(N<10)$ system in the realistic experiment. The origin of the multiple OMIT windows can be explained by the quantum interference effects between different energy level pathways, and the energy level configurations of the hybrid system consisted of $N$ cavities coupled with $N$ mechanical oscillators are presented in Fig.~\ref{fig3}. The excited pathway of the probe field is quantum interfering with different coupling pathways $G_{mj}(j=1,...,N)$ of the control field and the tunneling pathways $g_i(i=1,...,N)$. Therefore, the sum of the quantum interference pathways is $2N-1$ for $N$ cavities and $N$ mechanical oscillators. In addition, those pathways of the destructive quantum interference are formed via the optomechanical interaction and the tunneling, which lead to $2N\!-\!1$  transparency frequencies of the output field under the condition of $\varepsilon_T\!\approx\!0$ at extremum points.

\begin{figure}[t]
\centerline{\includegraphics[width=0.5\textwidth]{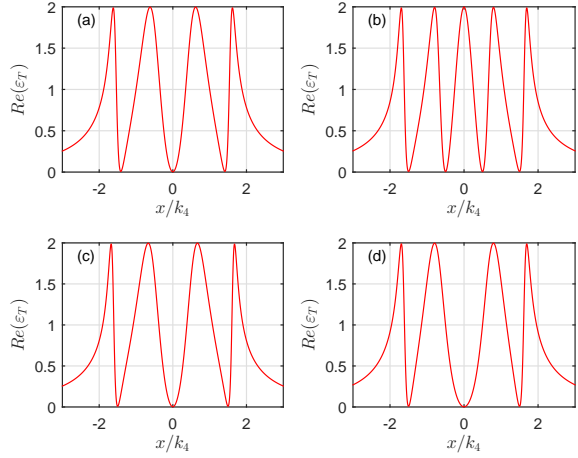}}
\caption{The absorption $Re(\varepsilon_T)$ as a function of $x/\kappa_{4}$ for four cavities. The subplot (a) corresponds to no mechanical oscillators coupled to cavities, the subplot (b) describes two mechanical oscillators coupled to cavity $1$ and $3$, respectively. The subplot (c) shows one mechanical oscillator coupled to cavity $2$, and the subplot (d) illustrates two mechanical oscillators coupled to the $2nd$ and $4th$ cavity, respectively.}
\label{fig4}
\end{figure}

To further explore the characteristics of the OMIT arising from the interaction of the mechanical oscillators, we plot the absorption $Re(\varepsilon_T)$ of the output field as a function of $x/\kappa_N$ for one and two coupled oscillators cases. The case without the mechanical oscillator coupling is also shown for comparison in Fig.~\ref{fig4}.
Due to the destructive interference between the pathways of the mechanical oscillator and the cavity field, the system will add a new transparency window if the first
cavity is coupled with a mechanical oscillator, which is shown in Figs.~\ref{fig2}(a) and ~\ref{fig4}(a). However, comparing Figs.~\ref{fig4}(b) with ~\ref{fig4}(a), it can be found that the third labelled mechanical oscillator just broaden the central absorptive peak. On the other hand, Figs.~\ref{fig4}(c) describes the coupling between the mechanical oscillator and $2nd$ cavity.
Fig.~\ref{fig4}(d) describes that the mechanical oscillators interact with $2nd$ and $4th$ cavity, respectively. Compared with Fig.~\ref{fig4}(a), it can be found that the even-labelled mechanical oscillators does not change the number of the transparency window for both case, only contributes to broaden the central absorptive dip compared to the case of without mechanical oscillator coupling.
Note that, although all the mechanical oscillators are identical, they can still lead to different quantum interference pathways.

The numerical calculation shows that, if one enlarges the numbers of the cavities and the odd- (even)-labelled mechanical oscillators, the results are similar with the ones mentioned above. In detail, for the odd-labelled case, the number of the transparency windows only adds one compared with the case of without mechanical oscillator coupling no matter how many mechanical oscillators are coupled with the cavities. And the increased odd-labelled mechanical oscillators only change slightly width of the central absorptive peak. While for the even-labelled ones, the increased oscillator only alter the width of the central absorptive peak or dip. These behaviors can be analyzed from Eq. (\ref{H13}). The equation of $\varepsilon_T\approx0$ has $N-1$ different roots without the coupled mechanical oscillator at the extremum points. For odd- (even)-labelled oscillator coupled with its cavity, $\varepsilon_T\approx0$ has at most $N$$(N-1$) different roots. Furthermore, when only odd- or even-labelled oscillators are coupled with the cavities, we also find that increasing the effective optomechanical rate $G_{mN}$, the central absorptive peak or dip will be remarkable broadened. As for the broadened central absorptive dip, the phenomenon of the destructive interference is weakened with the increase of the central absorptive dip of the output field, and the consequent EIT-Autler Townes splitting (ATS) crossover or ATS\cite{Autler} can occur. Due to the splitting of energy levels resulting from the strong field-driven interactions, identifying OMIT or EIT with ATS has been detailedly investigated in toroidal microcavity system\cite{BPeng} and the circuit circuit quantum electrodynamics system\cite{QCLiu,HCSun}.

\begin{figure}[htbp]
\begin{minipage}[t]{0.35\linewidth}
\centering
\includegraphics[height=9cm,width=7.5cm]{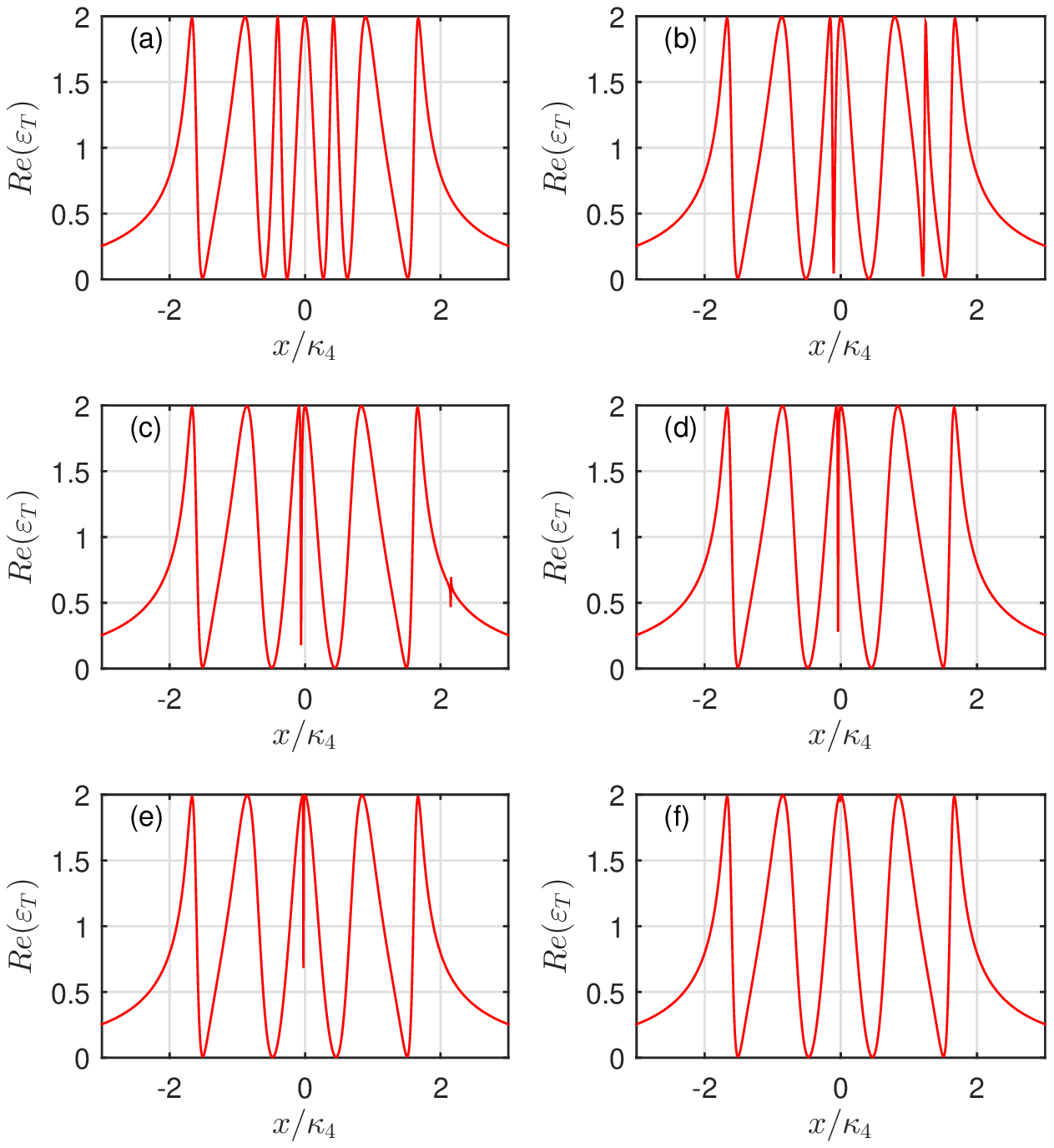}
\caption{ The absorption $Re(\varepsilon_T)$ as a function of $x/\kappa_4$. Figs. 5(a)-5(d) illustrate the cases of two Rydberg atoms trapped in $1st$ cavity coupled with the mechanical oscillator, and correspond to the DDI with $V(R)/g_{mj}=(0,2,4,6,10,30)$, respectively.}
\label{fig5}
\end{minipage}%
\hfill
\begin{minipage}[t]{0.5\linewidth}
\centering
\includegraphics[height=9cm,width=7.5cm]{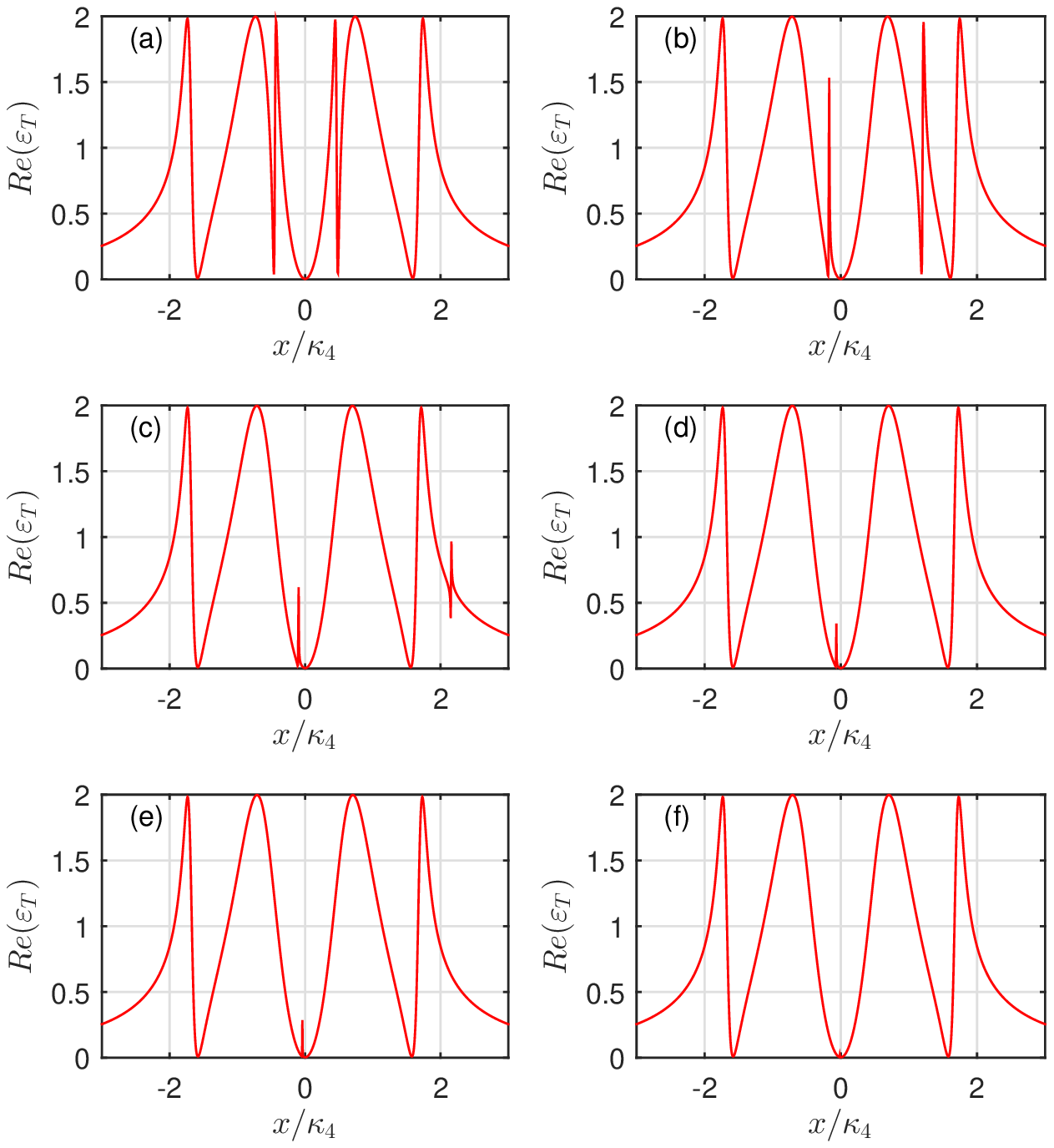}
\caption{ The absorption $Re(\varepsilon_T)$ as a function of $x/\kappa_4$. Figs. $6(a)$-$6(d)$ describe the cases of two Rydberg atoms trapped in $2nd$ cavity coupled with a mechanical oscillator, and correspond to DDI with different strengthes $V(R)/g_{mj}=(0,2,4,6,10,30)$, respectively.}
\label{fig6}
\end{minipage}
\end{figure}

\begin{figure}[htbp]
\begin{minipage}[t]{0.35\linewidth}
\centering
\includegraphics[height=5.5cm,width=7.5cm]{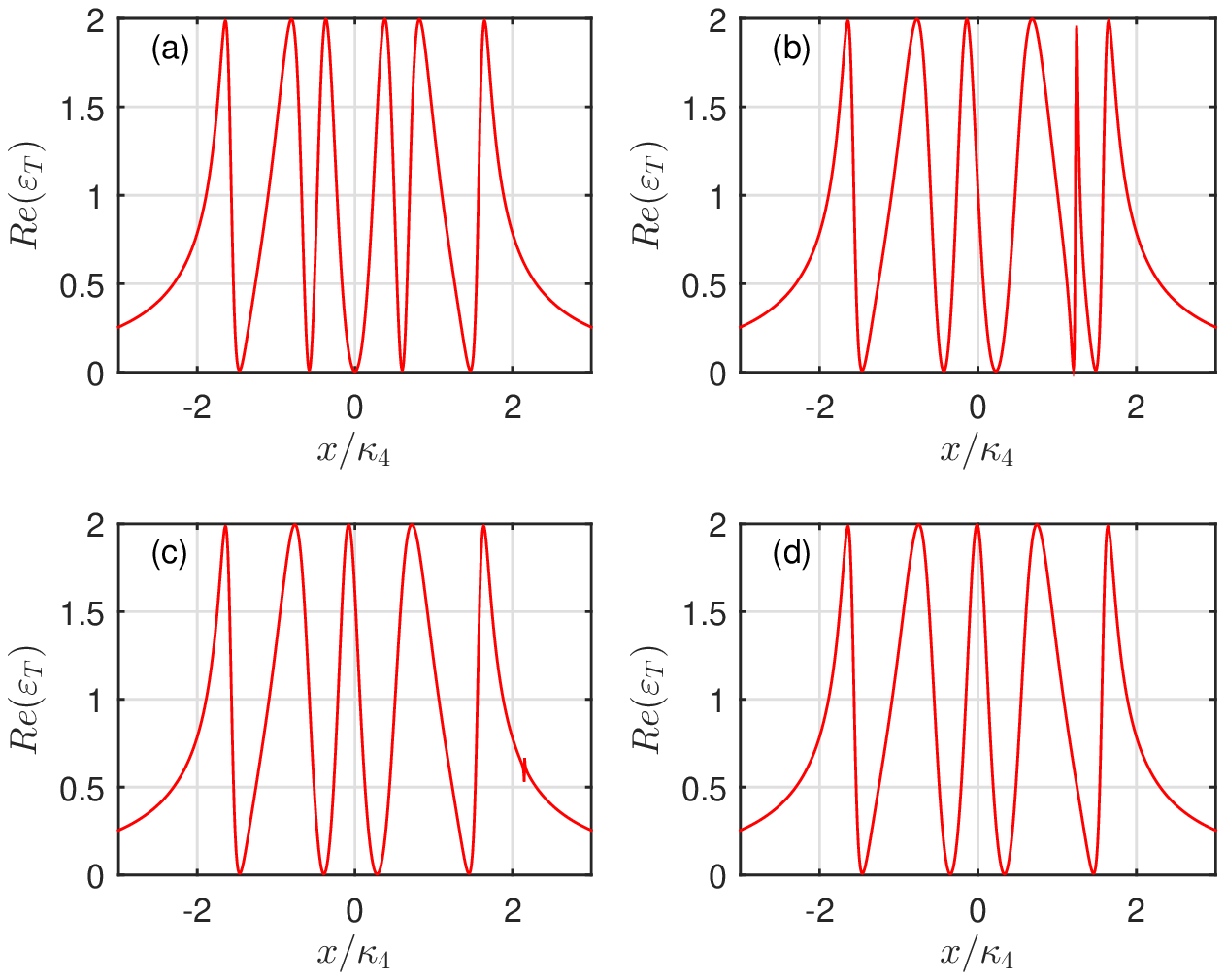}
\caption{ Real part $Re(\varepsilon_T)$ as a function of $x/\kappa_4$. Figs. $7(a)$-$7(d)$ illustrate the cases of two Rydberg atoms trapped in the $1st$ cavity and the mechanical oscillators do not couple with cavities, which correspond to DDI with $V(R)/g_{mj}=(0,2,4,30)$, respectively.}
\label{fig7}
\end{minipage}%
\hfill
\begin{minipage}[t]{0.5\linewidth}
\centering
\includegraphics[height=5.5cm,width=7.5cm]{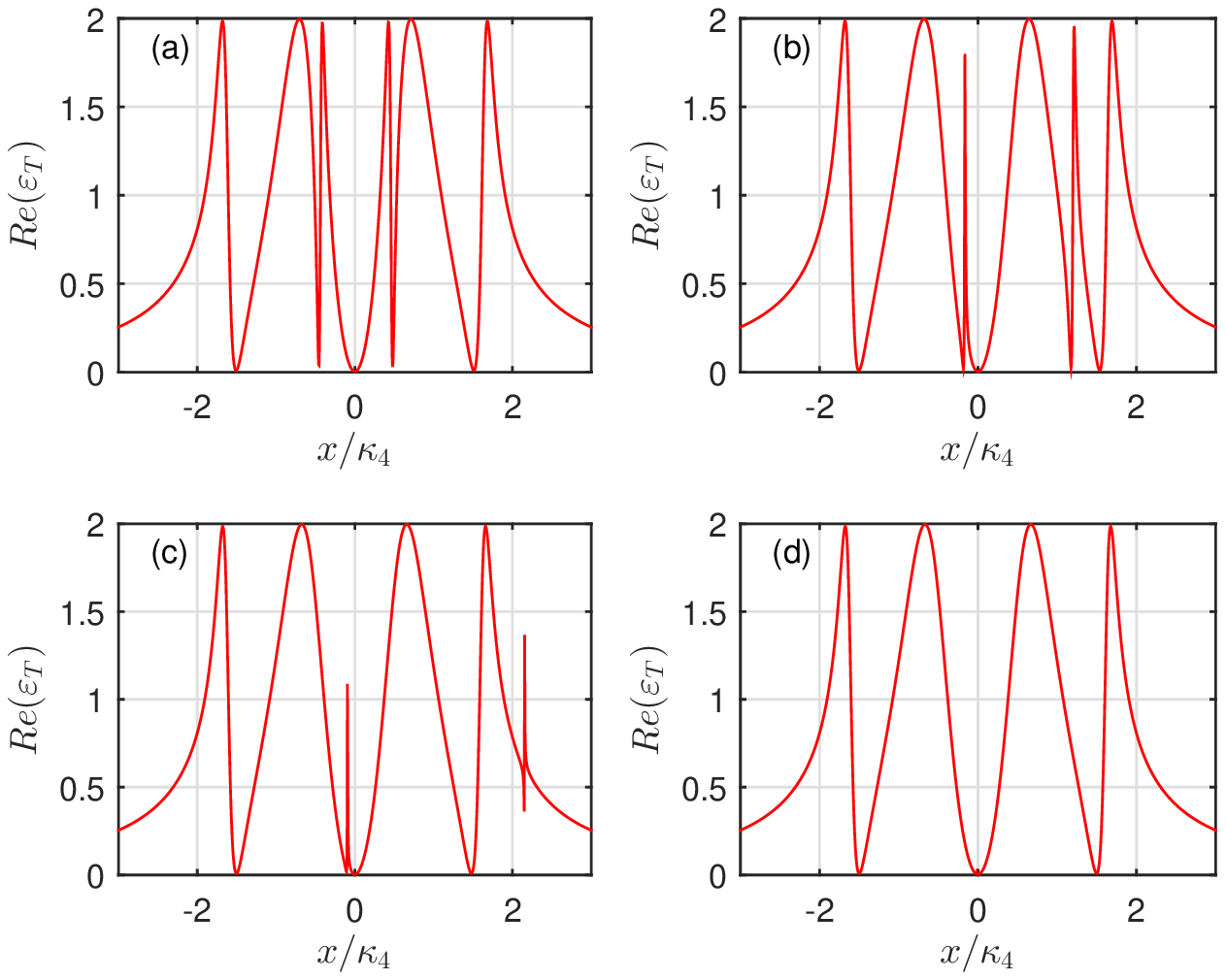}
\caption{ Real part $Re(\varepsilon_T)$ as a function of $x/\kappa_4$. Figs. $8(a)$-$8(d)$ describe the cases of two Rydberg atoms trapped in $2nd$ cavity  and the mechanical oscillators do not couple with cavities, which correspond to DDI with $V(R)/g_{mj}=(0,2,4,30)$, respectively.}
\label{fig8}
\end{minipage}
\end{figure}

\subsection{With Rydberg atoms.}
In the proceeding section, we have considered the variation of the multi-OMIT without the Rydberg aotms. Now, we shall investigate the multi-OMIT in the present system in which two Rydberg atoms are trapped in $ith(i=1,...,N)$ cavity and interact with the cavity field, and explore the effects of DDI on the OMIT.
The parameters  $\gamma_{rr}/g_{mj}=\gamma_{gr}/g_{mj}=\gamma_{ee}/g_{mj}=\gamma_{er}/g_{mj}=0.001$, $\Omega/g_{mj}=g/g_{mj}=1$. The other parameters are same as the ones in the previous section. In order to simplify the model and highlight the effect of the Rydberg atoms in the $ith$ cavity, we just only consider one mechanical oscillator which interacts with the $ith$ cavity as others do not affect the behavior of Rydberg atom directly in principle.

In general, the maximal DDI strength is of the order of gigahertz\cite{Li2005}. Figs.~\ref{fig5}(a)-(d) describe one mechanical oscillator interacts with the $1st$ cavity and the Rydberg atoms are also trapped in the same cavity with different DDI strength for four cavities. In Fig.~\ref{fig5}(a), when DDI strength is zero, one can find that two extra symmetric transparency windows (extra resonances) appear on both sides of the central absorptive peak compared to the case [See Fig. ~\ref{fig3}(a)] without Rydberg atom. One can also find that the positions of the two extra resonances move to the right with the increase of the DDI strength as shown in Figs.~\ref{fig5}(b)-(d). But the position of the left extra resonance moves slowly than the right one. In Fig.~\ref{fig6}, one mechanical oscillator and two Rydberg atoms coupled with $2nd$ cavity have been discussed. The variation tendencies of two extra resonances are the same as the ones in Fig.~\ref{fig5}. However, the widths, the positions and the amplitudes of two extra resonances are different. When the Rydberg atoms are trapped in $3rd$ and $4th$ cavities, numerical results also show that same variation tendencies of two extra resonances can be obtained in Figs.~\ref{fig5} and ~\ref{fig6}, respectively. But the widths and the amplitudes of the two extra resonances have little difference.

In addition, the amplitudes of two extra resonances become smaller and experience Fano resonance with the increase of DDI strength. When DDI strength increases, the left extra resonance gets close to the central absorptive dip and then both extra resonances die out. Compared Fig.~\ref{fig5}(d) (~\ref{fig6}(d)) with Fig.~\ref{fig2}(a) (~\ref{fig4}(c)), we can find that the DDI only impacts on the width of central absorptive dip or peak when the DDI strength is large. Therefore, the large DDI strength of Rydberg atoms has slight influence on the output field. On the other hand, from Eq.~(\ref{H13}), one can find that the DDI strength can adjust the effective detunings $x_{gr}$ and $x_{er}$, which makes the OMIT be sensitive to the DDI strength. As we all know, with the change of effective detuning, the extra OMIT windows can move and become a Fano line shape\cite{Qu}.Then the extra narrow OMIT window, a analogue to EIT, evolves into a Fano resonance in the output field of the hybrid optomechanical system with the increase of DDI strength between two Rydberg atoms.

In Figs.~\ref{fig5} and ~\ref{fig6}, we discuss the influences of DDI strength and the mechanical oscillator coupling strength in the absorption of the output field.
But we only consider the factor of DDI strength in Figs.~\ref{fig7} and ~\ref{fig8}. Compared Figs.~\ref{fig5} (\ref{fig6}) with Figs.~\ref{fig7} (\ref{fig8}), it can be found that the same behavior of the output filed appears except the slight differences in the position and width of the transparency windows compared with the cases of wihtout mechanical oscillator coupling. In detail, there are two additional transparency windows for weak DDI strength. When $V(R)$ becomes more and more greater, two extra windows move and become Fano resonance till the right extra resonance of the absorption profile disappears gradually and the left extra resonance approaches the central absorptive peak. Note that, the system reduces to a coupled cavity system assisted a two-level atom in the large range DDI strength\cite{Sohail}. Because the influence of the coupled Rydberg atoms resembles a mechanical oscillator as mentioned above. If the positions of the atoms is different, the different numbers of the transparency window appear as shown in Figs. ~\ref{fig7}(d) and ~\ref{fig8}(d). This result may be explained as follows. When DDI strength between Rydberg atoms is relatively weak, it is obvious that the second excited Rydberg atom does not shift the level of the first one. The system is regarded as a coupled cavity interacted with both a mechanical resonator and ladder-type Rydberg atoms.
Due to the transitions $|g\rangle \leftrightarrow |e\rangle$ and  $|e\rangle \leftrightarrow |r\rangle$ of the Rydberg atom in the hybrid system, additional interference pathways appear. Therefore, two additional OMIT windows in the absorption profile are observed. With the increase of DDI strength, the Rydberg blockade suppresses the excitation of the first atom and makes the OMIT condition be no longer fulfilled for the first atom. Then the first atom acts as a two-level atom which couples resonantly to the probe field.

\section*{{\protect\LARGE \textbf{Conclusion and Discussion}}}

In summary, we have studied the OMIT of the MCOS. For the case without  Rydberg atoms trapped in the cavity, the MCOS system has been demonstrated the generation of $2N\!-\!1(N<10)$ OMIT windows for the output field, when $N$ cavities interact with $N$ mechanical oscillators, respectively. But the odd- and even-labelled oscillators will lead to different effects, if the odd-labelled oscillators are presented, only one extra OMIT emerges in the absorption profile by the quantum interference. In contrast, the increased even-labelled mechanical oscillators just broaden the central absorptive dip or peak. Under these circumstances, the corresponding transparency window can change from OMIT to ATS by increasing the effective optomechanical rate.
On the other hand, when two Rydberg atoms are trapped in the $ith$ cavity with weak DDI and the cavity is coupled with a mechanical oscillator, two extra OMIT windows can be observed. In addition, two extra OMIT windows would gradually move to the far off-resonance regime with the DDI strength increasing. The right extra resonance move faster with the increase of the DDI strength. But the right one vanishes with great DDI strength. Furthermore, Fano resonances also appears with the changes of DDI strength.

In experiment,
one possible scheme is the toroidal microcavity-tapered optical fiber system coupled with Rydberg atoms. Firstly, the effect of OMIT in a single optical nanofiber-based photonic crystal optomechanical cavity has been engineered in the experiments\cite{Huangjy,BPeng}. Further, a two-color optical dipole trap has also come true by using the red- and blue-detuned evanescent light fields near the optical nanofiber. This method can allow the Rydberg atoms to be prepared at a few hundred nanometers from the nanofiber surface and coupled with the ith photonic crystal cavity\cite{Vetsch,Goban}. And a series of nanofibers acted as a 1D coupled cavity array has been realized experimentally\cite{Notomi}, which is extended to lattices of coupled resonators with
Rydberg atoms\cite{ZhangY}. Therefore, combined with the above experiments, the multi-cavity optomechanical system with two Rydberg atoms trapped in one cavity may be realizable with the present-day or near-term technology.

\subsection*{Acknowledgments}
 This work was supported by the National Natural Science Foundation of China under grants Nos. 11274148 and 11434015, the National Key R$\&$D Program of China
under grants Nos. 2016YFA0301500, and SPRPCAS under grants No. XDB01020300, XDB21030300.

\subsection*{Author contributions}
 J.-L.M., L.T., Q.L., H.-Q.G., and W.-M.L. conceived the idea. J.-L.M. performed the theoretical as well as the
numerical calculations. J.-L.M. and L.T. interpreted physics and wrote the manuscript. All of the authors reviewed the manuscript.

\subsection*{Additional Information}
\textbf{ Competing financial interests:} The authors declare that they have no
competing interests.

\end{document}